\documentclass[sigconf]{acmart}
\usepackage{bm}
\usepackage{multirow}

\AtBeginDocument{%
  \providecommand\BibTeX{{%
    \normalfont B\kern-0.5em{\scshape i\kern-0.25em b}\kern-0.8em\TeX}}}

\setcopyright{acmcopyright}
\copyrightyear{2022}
\acmYear{2022}
\acmDOI{XXXXXXX.XXXXXXX}

\acmConference[CIKM '22]{the 31st ACM International Conference on Information and Knowledge Management (CIKM '22)}{October 17--22,
  2022}{Atlanta, Georgia}
\acmPrice{15.00}
\acmISBN{978-1-4503-XXXX-X/18/06}

\begin{document}

\title{KQGC: Knowledge Graph Embedding with Smoothing Effects of Graph Convolutions for Recommendation}

\author{
Daisuke Kikuta$^{1*}$,
Toyotaro Suzumura$^{1, 3}$,
Md Mostafizur Rahman$^{1}$, 
Yu Hirate$^{1}$, \\
Satyen Abrol$^{2}$, 
Manoj Kondapaka$^{2}$,
Takuma Ebisu$^{1}$,
Pablo Loyola$^{1}$
}
\thanks{$^*$The author was with Rakuten Institute of Technology when the work was done}
\affiliation{%
 \vspace{2.5mm}$^{1}$Rakuten Institute of Technology, Rakuten Group, Inc., Tokyo,  Japan
}
\affiliation{%
 $^{2}$Rakuten Institute of Technology, Rakuten Group, Inc., Bengaluru, India
}
\affiliation{%
 $^{3}$The University of Tokyo, Tokyo, Japan
}
\affiliation{%
  kiku.btl.a@gmail.com, suzumura@acm.org,\\
  \{mdmostafizu.a.rahman, yu.hirate, satyen.abrol, kondapaka.manoj, takuma.ebisu, pablo.a.loyola\}@rakuten.com
}

\renewcommand{\shortauthors}{Kikuta, et al.}

\begin{abstract}
Leveraging graphs on recommender systems has gained popularity with the development of graph representation learning (GRL). In particular, knowledge graph embedding (KGE) and graph neural networks (GNNs) are representative GRL approaches, which have achieved the state-of-the-art performance on several recommendation tasks. Furthermore, combination of KGE and GNNs (KG-GNNs) has been explored  and found effective in many academic literatures.  \par
One of the main characteristics of GNNs is their ability to retain structural properties among neighbors in the resulting dense representation, 
which is usually coined as smoothing. The smoothing is specially desired in the presence of homophilic graphs, such as the ones we find on recommender systems.
In this paper, we propose a new model for recommender systems named Knowledge Query-based Graph Convolution (KQGC). In contrast to exisiting KG-GNNs, KQGC focuses on the smoothing, and leverages a simple linear graph convolution for smoothing KGE.
A pre-trained KGE is fed into KQGC, and it is smoothed by aggregating neighbor \textit{knowledge queries}, which allow entity-embeddings to be aligned on appropriate vector points for smoothing KGE effectively.
We apply the proposed KQGC to a recommendation task that aims prospective users for specific products. Extensive experiments on a real E-commerce dataset demonstrate the effectiveness of KQGC. 
\end{abstract}



\maketitle

\section{Introduction}
Internet services, including e-commerce and streaming,  have been growing up rapidly in recent years. Recommender Systems (RecSys) are indispensable technologies for enhancing user experience of these services. 
Therefore, how to fully leverage rich sources of data and improve RecSys have been one of the most critical challenges for companies in the field.
RecSys have been widely studied in both research and industry, and  various approaches have been proposed over the years.
Most of them have relied on collaborative filtering (CF), which infers user's latent preferences based on the  co-occurrence in historical interactions. Traditional CF-based methods, such as user-based model \cite{UserBasedCF1, UserBasedCF2}, item-based model \cite{ItemBasedCF1, ItemBasedCF2, Item_CF} and matrix factorization \cite{MF}, generate user's and item's features from an interaction matrix. However, they cannot incorporate side information, resulting in poor performance in the cold start problem \cite{KGAT, KGCN}.\par
Graph representation learning-based approaches have 
gained popularity in recent years due to the ability to handle with both high-order relations in historical interactions and side information. In particular, Graph Neural Networks (GNNs) have shown prominent performance in RecSys based on their ability of modelling complex settings and scalability. User-item bipartite graphs have been employed for GNNs traditionally \cite{PinSAGE, GCMC, SpectralCF, NGCF, MCCF}, while utilization of knowledge graphs (KGs) have been also considered lately \cite{KGCN, KGAT, KGIN}.
Leveraging KGs has two main advantages. Firstly, it can handle attributes relationships (e.g., hierarchical dependency of product genres in e-commerce). Secondly, extracting attributes as entities (nodes) in a KG allows two nodes having the same attributes to be connected through the attribute entities. As a result, GNNs can capture high-order relations by sharing attributes more efficiently \cite{KGAT}. The representative models combining KG and GNNs (KG-GNNs) are KGCN \cite{KGCN} and KGAT \cite{KGAT}. KGCN is a meta-path based GNN, which generates embeddings of entities and relations in the KG with knowledge-aware attention-based GNN layers. 
KGAT consists of two layers: Knowledge Graph Embedding (KGE) and GNN layers, and it generates node embeddings by applying GNN layers to the KGE obtained from TransR \cite{TransR}. Here, the two layers are trained simultaneously.\par
We argue that homophily holds in graphs that emerge in recommendation tasks, and aggregation (smoothing) is the most effective function of GNNs there as Simple Graph Convolution \cite {SGC} has shown. In this paper, we propose Knowledge Query-based Graph Convolution (KQGC), which enhances KGE with smoothing. KQGC relies on the simple principle of smoothing, and it is a KG-GNN model that has different perspective to KGCN and KGAT. Therefore, there are three distinctions between existing KG-GNNs and KQGC: 1) KQGC consists of both KGE model and graph convolution model. KGE is pre-trained, and subsequently the KGE is fed into graph convolution model. i.e., the two models are trained separately in contrast to KGAT. 2) KGCN and KGAT aggregates only neighbor entity embeddings, while KQGC aggregates knowledge queries, which are sums of entity and relation embeddings. That can alleviate irrelevant smoothing. 3) KGCN and KGAT apply nonlinear function when updating embeddings, while KGCN does not. As a results, property of the pre-trained KGE is preserved.\par
Rakuten group has one of the largest E-commerce platform in the world.  In particular, AIris Target Prospecting (AIris TP), which is a framework that finds prospective users for Rakuten business clients, has been playing an important role in this E-commerce platform. Companies like Facebook runs one of the world’s most powerful advertising platforms, which is called "Facbook Lookalike Audience". Rakuten AIris TP serves the same purpose to its clients. Therefore, we aim to improve the performance of AIris TP in this paper. We apply KQGC to the Airis TP, and evaluate the effectiveness.\par
The main contributions of our work are organized as follows:
\begin{itemize}
    \item We propose a new model combining KGE and graph convolution called KQGC, which enhances KGE with smoothing effects of graph convolution. It provides a new perspective of combining KGE and GNNs.
    \item Experimentally, we show the effectiveness of KQGC and its smoothing effects on real E-commerce dataset.
\end{itemize}

\section{Related Work}
Graph representation learning (GRL) for RecSys is roughly categorized into three approaches: knowledge graph embedding (KGE), graph neural networks (GNNs) and combination of the two.
In this section, we describe each of them and clarify relation between them and the proposed model.

\subsection{\textbf{Knowledge Graph Embedding}}
Knowledge graph (KG) is a directed heterogeneous graph that represents a fact, where nodes and edges represent entities and relations between two entities, respectively. It is formulated as a set of triples $\{(h,r,t)|h,t\in\mathcal{E}, r\in\mathcal{R}\}$, where $h$, $t$, $r$ are heads, tails and relations. For instance, the fact that shoes A is sold by Shop B is represented as a triple $(Shoes A, isSoldBy, Shop B)$. The main goal of KGE is to map enitities and relations in KGs to embeddings (vectors).\par
There are two representative categories of KGE models: bilinear (semantic matching) model and translation-based model \cite{KGE_review}. 
Bilinear model is a tensor factorization model, which generates embeddings of entities and relation matrices that can reconstruct relational adjacency matrix of the KG. RESCAL \cite{RESCAL}, DistMult \cite{DistMult} and ComplEX \cite{ComplEX} are representative bilinear models.
In translation-based model, embeddings of entities and relations are initialized with random values, then they are updated so that distance between embeddings of positive triples (e.g., $||\bm{e}_h+\bm{e}_r-\bm{e}_t||_{1,2}$ in TransE) is minimized.
TransE \cite{TransE} is the first translation-based model, and there are several variants of it such as TransR \cite{TransR} and TransD \cite{TransD}.\par
In the context of RecSys, recommended items or users are determined by evaluating similarity score between user- and item-embeddings obtained by KGE models. More specially, in the case that we recommend $k$ items to a user, top-$k$ items that have high score (e.g., inner product between the user and items with purchase relation $(\bm{e}_u+\bm{e}_{purchase})^T\bm{e}_{i}$) are selected as the recommended items.
Our model employs translation-based models in the pre-training phase, and subsequently smooths the pre-trained KGE with graph convolution. Consequently, our model strengthens the homophily of KGE.

\subsection{\textbf{Graph Neural Networks}}
Graph Neural Networks (GNNs) are neural network models for graphs, which encode node/edge embeddings by leveraging both node/edge features and graph structure.
There are three types of GNNs: recurrent GNNs, spectral-based convolutional GNNs and spatial-based convolutional GNNs (spatial-based ConvGNNs) \cite{GNN_overview}. In particular, many variants of spatial-based ConvGNNs have been proposed actively in recent years \cite{GCN, GraphSAGE, GAT, RGCN, HGT} due to the flexibility of modeling and scalability.
Generally, GNNs are formulated with the message passing (MP) framework \cite{MPNN}, which consists of two steps: aggregation and update steps. Firstly, the aggregation function aggregates neighbor node features of destination nodes, where we term updated nodes destination nodes. The update function then updates the destination node features by combining the aggregated neighbor node features with themselves. Here, neural networks are employed in both the aggregation and update functions usually. \par
GNNs have shown their remarkable capabilities for RCs as well as classification tasks over the years. User-item bipartite graphs, which are constructed with user-item interactions, are utilized as the input graphs in the literature. Furthermore, user demography features and item attributes are utilized as node features in contrast to KGE which treats the features as entities (nodes). Aggregating neighbor node features is regarded as collecting collaborative signals. Hence GNN-based approaches for RecSys are the CF-based models with side information, that possess various expansions for complicated modeling and large scale applications. \par
On the other hand, a paper has argued that nonlinearity in GNNs is not necessary needed for several tasks. The authors have proposed Simple Graph Convolution (SGC) \cite{SGC}, which is a graph convolution without activation function, and shown its competitive performance compared to the state of the art GNNs. The simplification improves interpretability of the model and reduce computational complexity.
We believe that the fact holds in tasks related to homophily and recommendation tasks are involved in there. In this paper, we employ a new graph convolution without activation function, which allows the property of the traslation-based models to be preserved. 
\subsection{\textbf{Combination of KGE and GNNs}}
Two different GRL approaches are described in the above, while combination of the two has been also studied in recent years. 
In \cite{KGAT}, the authors introduce collaborative knowledge graph (CKG), which merges a user-item bipartite graph and a KG of items, and apply KGAT to it. KGAT is a model that combines KGE and GNNs. Firstly, it generates KGE with TransR \cite{TransR}, and subsequently the KGE is fed into knowledge-aware attention-based GNN layers. TransR and GNN layers are trained simultaneously with the loss function of TransR and a pair-wise loss function.
KGCN \cite{KGCN} is a mata-path based GNN model, which consists of only GNN layers designed for KGs.
In KGCN, embeddings of entities and relations are initialized with random values as with normal KGE models, then the embeddings are updated with knowledge-aware attention-based GNN layers. 
Nodes that have the same attributes are connected each other in KGs.
That allows GNNs to capture high-order relations explicitly, resulting in enhancing effectiveness of aggregation of GNNs. In fact, they reaches the state of the art performance in several recommendation tasks.\par
However, the two models aggregates only entity embeddings in GNN layers, and that possibly incurs irrelevant smoothing (described in Section \ref{subsec:KQGC}). 
To alleviate this issue, the proposed model aggregates knowledge queries, which consists of both entity and relation embeddings. The proposed framework has two components: KGE and a linear graph convolution, and they are trained separately in contrast to KGAT.   

\section{Methodology}
\begin{figure*}[htbp]
    \begin{center}
    \includegraphics[width=\linewidth]{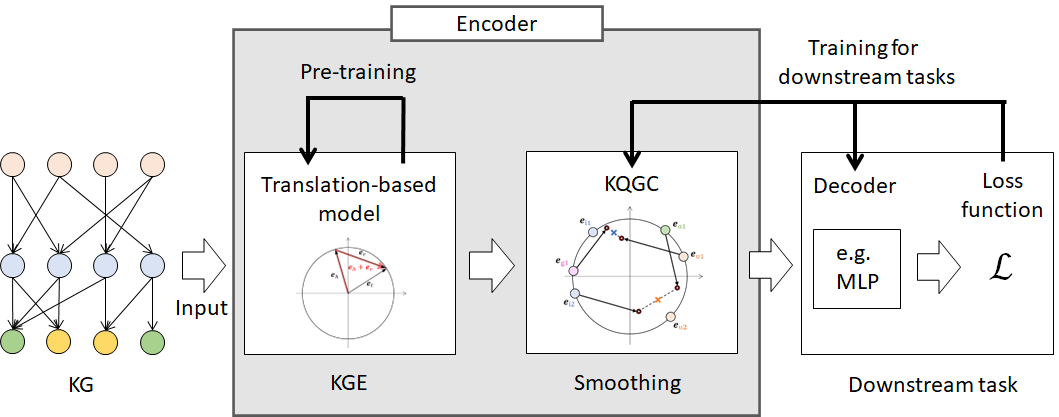}
    \caption{Framework of KQGC}
    \label{fig:KQGNN_architecutre}
    \end{center}
\end{figure*}

We now introduce the proposed Knowledge Query-based Graph Convolution (KQGC), which improves pre-trained KGE with smoothing effect of graph convolution.
Figure \ref{fig:KQGNN_architecutre} shows the overall framework, which consists of two models: translation-based KGE model and KQGC. 
The whole procedure for applying KQGC to downstream tasks is as follows: 
1) a knowledge graph (KG) is fed to the knowledge graph embedding (KGE) model, then KGE for all the entities and relations are trained in unsupervised learning;
2) the pre-trained KGE is input to KQGC as initial node/edge features, and the KQGC is trained with downstream tasks.
Note that the KGE is fixed when KQGC is trained, i.e., the two models are trained separately.
We hereafter describe the details of the two models (Section \ref{subsec:KGE_model} and \ref{subsec:KQGC}), and then introduce a use case of KQGC in Rakuten group (Section \ref{subsec:usecase}).

\subsection{Translation-based Knowledge Graph Embedding model}
\label{subsec:KGE_model}
\begin{figure}[htbp]
    \begin{center}
    \includegraphics[width=150pt]{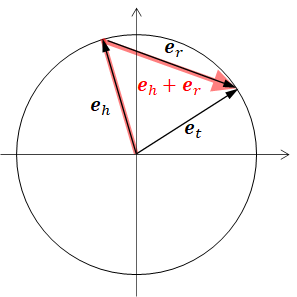}
    \caption{A relation between embeddings of positive triples obtained by TransE; For simplicity, a 2D case is depicted. As each embedding is normalized, it is mapped on the unit circle.}
    \label{fig:TransE}
    \end{center}
\end{figure}

In this work, we employ TransE as the translation-based KGE model. In TransE, the embeddings of all the entities and relations are initialized with random values, then they are updated by the gradient descent method based on the max-margin loss function $\mathcal{L}_{\textrm{KGE}}$.
\begin{align}
\label{eq:loss_kge}
    \mathcal{L}_{\textrm{KGE}} &= \sum_{(h, r, t)\in\mathcal{E}}\sum_{(h', r, t')\in\mathcal{E}^{-1}}\left[\gamma + f(h,r,t) - f(h',r',t')\right]_+,
\end{align}
where $(h,r,t)$ is positive triples, which actually exist in the KG,  $(h',r,t')$ is broken triples, $\gamma$ is the margin, $f(h,r,t)$ is the score function, and $[\cdot]_+=max(0, \cdot)$. The broken triples are generated by replacing either head- or tail-entity of the positive triples with an entity randomly chosen. The score function of TransE is as bellow. 
\begin{equation}
    \label{eq:transE}
    f(h,r,t) = \|\bm{e}_h + \bm{e}_r - \bm{e}_t \|_{1,2},
\end{equation}
where $\bm{e}_h$, $\bm{e}_r$ and $\bm{e}_t$, are the embeddings of head entities, relations and tail entities, respectively. In Eq. (\ref{eq:transE}), either L1 or L2 norm is usually used, we employ L1 norm. Minimizing Eq. (\ref{eq:loss_kge}) with this score function (Eq.(\ref{eq:transE})) leads a relation between embeddings of positive triples $\bm{e}_h + \bm{e}_r\approx\bm{e}_t $ in the same vector space (Figure \ref{fig:TransE}). \par
From the other perspective, sum of a head and relation embeddings (the red arrow in Figure \ref{fig:TransE}) reaches a near point to the related tail embeddings, therefore it can be regarded as a query to visit tail entities related to the head entity with the relation. In this paper, we term it \textit{knowledge query}. KQGC, which is described in the next section, aggregates the knowledge queries obtained by TransE for aligning neighbor node embeddings in the vector space.

\subsection{Knowledge Query-based Graph Convolution}
\label{subsec:KQGC}
After training of KGE is completed, the trained KGE is fed into KQGC and smoothed by it. 
KQGC follows the MP framework and consists of two steps: aggregation and update. \par
\begin{figure*}[htbp]
    \begin{center}
    \includegraphics[width=\linewidth]{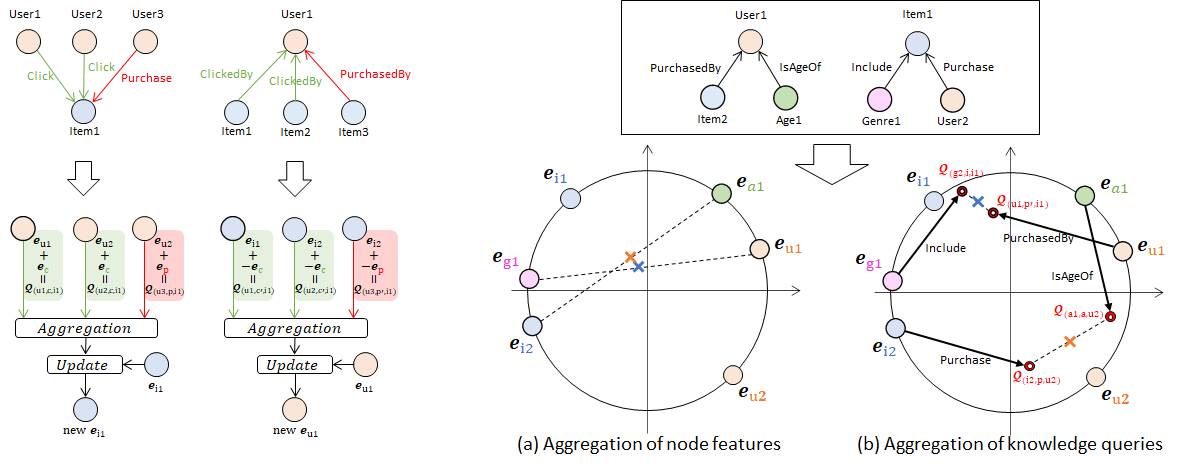}
    \caption{Aggregation of knowledge queries in KQGC}
    \label{fig:query_aggregation}
    \end{center}
\end{figure*}

In terms of aggregation, KQGC aggregates \textit{knowledge queries} instead of neighbor node features. Aggregation of knowledge queries is depicted in Figure \ref {fig:query_aggregation}. The knowledge query from a source node to the destination node with a relation, which is aggregated when updating the destination node feature, is defined as
\begin{equation}
    \mathcal{Q}_{(src, r, dst)} = \bm{e}_{src} + \bm{e}_{r},
\end{equation}
where $\bm{e}_{src}$ and $\bm{e}_{r}$ are embeddings of the source node and the relation, and the triple $(src, r, dst)$ exists in the KG. 
In the case that the KG is a directed graph and we have only embedding of one direction relation, we obtain that of the reversed relation by multiplying the embedding by -1. 
One of the main efficacy of aggregating knowledge queries is alignment of node embeddings in the vector space, which is shown in the right-hand side of Figure \ref{fig:query_aggregation}.
We think of the case that updates embeddings of user A and item A as an example. For simplicity, let the aggregation be simple mean operation here. Assuming that each of the entity and relation embeddings is mapped on the unit circle as shown in Figure \ref{fig:query_aggregation}, the aggregated (averaged) neighbor node features for both user A and item A end up being similar (Figure \ref{fig:query_aggregation} (a)), even though they do not have any nodes shared in common. On the other hand, the aggregated knowledge queries are aligned to near points to the embeddings of the destination nodes, resulting in preservation of similarity obtained from KGE model (Figure \ref{fig:query_aggregation} (b)). \par
In the update phase, KQGC combines the aggregated knowledge queries with the target node embedding using linear transformation. We do not leverage non-linear function in this phase in order to preserve the property of translation-based model when multiple layers of KQGC are stacked. SGC has shown adequate capability of graph convolution without nonlinear function for several tasks related to homophilicity. Intuitively recommendation task is included to the tasks, and we believe that absence of nonlinear function is not critical. 
We provide formulations of aggregation and update in below.

\subsubsection{Aggregator}
Here we propose mean- and attention aggregators. We redefine notation of knowledge queries for multiple KQGC layers, as follows.
\begin{equation}
    \mathcal{Q}^{l-1}_{(u,r,v)} = \bm{h}^{l-1}_{u} + \bm{h}^{l-1}_{r},
\end{equation}
where $l$ indicates $l$-th layer, $\mathcal{Q}^{l}_{urv}$ is the $l$-th query from source $u$ to the destination $v$ with relation $r$, 
$\bm{h}^{l}_{u}$ and $\bm{h}^{l}_{r}$ are the $l$-th embeddings of the source and relation, respectively. 
KGE obtained from TransE is utilized as initial embeddings of all the nodes and relations, i.e., $\bm{h}^{0}_{u}=\bm{e}_u$, $\bm{h}^{0}_{r}=\bm{e}_{r}$. For the second and subsequent ones, node and relation embeddings outputted from the previous layer are input to the next layer. The two types of aggregators are formulated as follows.
\subsubsection*{\textit{Mean aggregator}}
This aggregator simply takes the average of neighbor knowledge queries, which is formulated as,
\begin{align}
    \label{eq:mean_agg}
    \bm{m}_{\mathcal{Q}}^l &= \textrm{MEAN}\left(\left\{\mathcal{Q}^{l-1}_{(u, r, v)}, u\in\mathcal{N}(v), r \in \mathcal{R}(u, v) \right\}\right),
\end{align}
where $\bm{m}_{\mathcal{Q}}^l$ is the $l$-th message of knowledge queries, $\mathcal{N}(v)$ is the set of neighbor nodes of node $v$, $\mathcal{R}(u,v)$ is the set of relations between $u$ and $v$.
\subsubsection*{Attention aggregator}
Rather than aggregating neighbor knowledge queries equally, attention aggregator considers the importance of each knowledge queries and aggregates them with different weights. In this paper, we provide two types of attention aggregators, which are formulated as,
\begin{align}
    \label{eq:attn_agg}
    \bm{m}^l_{\mathcal{Q}} &= \sum_{u\in\mathcal{N}(v)}\alpha_{(u,r,v)}\mathcal{Q}^{l-1}_{(u,r,v)}, \\
    \label{eq:attn}
     \alpha_{(u,r,v)} &= \frac{e_{(u,r,v)}}{\sum_{k\in\mathcal{N}(v)}e_{(k,r,v)}},
\end{align}
\begin{align}
    \label{eq:attn1}
    (Attention1)\ \ \ e_{(u,r,v)} &= \left(\mathcal{Q}^{l-1}_{(u,r,v)}\right)^T\bm{h}_{v}^{l-1}, \\ 
    \label{eq:attn2}
    (Attention2)\ \ \ e_{(u,r,v)} &= \textrm{LeakyReLU}\left(\bm{a}^T\left(\mathcal{Q}^{l-1}_{(u,r,v)}\parallel\bm{h}^{l-1}_{v}\right)\right),
\end{align}
where $\alpha_{(u,r,v)}$ is normalized attention coefficient, $\bm{a}\in\mathbb{R}^{2H}$ is the trainable parameter, $H$ is dimension of the embeddings, and $\parallel$ denotes concatenation.
The first one (Eq. (\ref{eq:attn1})) employs inner product between the knowledge query and the destination node to calculate attention coefficients, which is the same as the attention mechanism of KGCN. Knowledge queries that are closer to the destination node in the vector space are aggregated with higher weights in this attention mechanism. The second one (Eq.(\ref{eq:attn2})) has trainable parameters and they allow us to automatically design how to aggregate knowledge queries with appropriate attention coefficients, based on the loss function of the downstream tasks. The formulation is inspired by the self-masked attention mechanism of GAT.
\subsubsection{Update} After aggregating knowledge queries, new embeddings of all the nodes and relations are obtained by combining the destination node embedding with the aggregated knowledge queries. We employ the following update rule.
\begin{align}
    \bm{h}_v^{l} &= W^l(\bm{h}_v^{l-1}+\bm{m}^l_{\mathcal{Q}}) + \bm{b}^l, \\
    \bm{r}^{l} &= W^l\bm{r}^{l-1} + \bm{b}^l,
\end{align}
where $\bm{h}_v^{l}$ and $\bm{r}^{l}$ are the updated embeddings of the target nodes $v$ and all the relations, which are inputs of the next layer, $W^l$ and $\bm{b}^l$ are the $l$-th trainable weight parameters. This update rule simply sums up the destination node embedding and the aggregated knowledge queries, and then map them to new vector space with linear transformation.
The transformation is also applied to the relation embeddings so that translation between entites and relations can be constructed in the new vector space as well. Therefore we do not employ nonlinear function in this formulation. 

\subsection{Applications of KQGC in Recommender Systems of Rakuten Group}
\label{subsec:usecase}
\begin{figure}[htbp]
    \begin{center}
    \includegraphics[width=\linewidth]{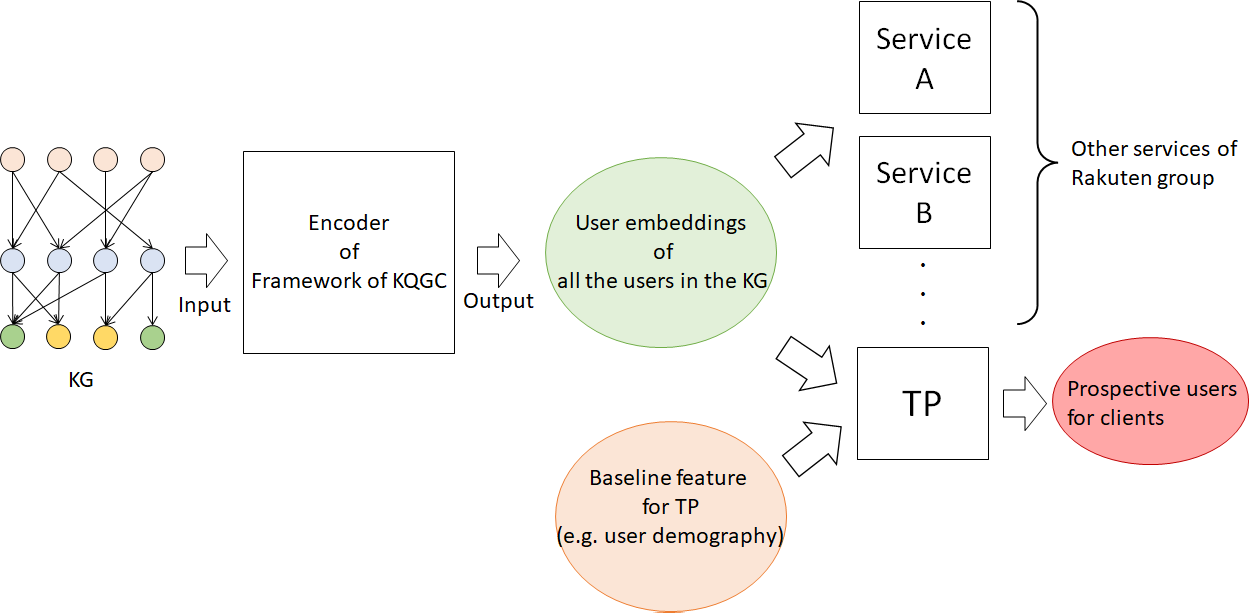}
    \caption{A framework for applying KQGC to several Rakuten services}
    \label{fig:use_case}
    \end{center}
\end{figure}

Rakuten group has various services and each service employs different model, therefore it is impossible to replace all the models with end-to-end fashion of KQGC. Instead, we attempt to improve the services by generating general user embeddings that can be used across various Rakuten services.
A framework for applying KQGC to various Rakuten services is depicted in Figure \ref{fig:use_case}. In this framework, KQGC and models in different services are completely separated for the portability. User embeddings obtained from KQGC are utilized in common as input features of each model with the baseline features. Types of the baseline features depend on types of services, which include user demographic features or user's purchase history. \par
In this paper, we focus on a Rakuten service called Target Prospecting (TP), and evaluate KQGC on the task of TP. The task of TP is to find prospective customers for clients (e.g., shops or brands) and prepare a list of top k active Rakuten users. The definition of active users is: the users who buy from Rakuten Ichiba regularly within average buying period. Currently, we have more than 40 millions active users. This paper employs XGBoost, which is a binary classifiers, as the service-side model. i.e., user embeddings obtained from KQGC and the AIris baseline features (refer to Section \ref{subsec:experimental_setup}) are input to the binary classifier, and it classifies prospective customers with true/false. We hereafter describe how to construct a knowledge graph for KQGC and how to train KQGC for our AIris TP task.
\subsubsection{How to construct knowledge graphs}
\label{subsubsec:how_to_construct_graph}
Rakuten group has the largest E-commerce platform called Rakuten Ichiba, which stores rich data of user behavior and item attributes in there. Purchase behavior of users is intuitively informative for not only predicting perspective users for clients but also capturing user preferences, which can be generally used for other services.
Hence we leverage the data on Rakuten Ichiba\footnote{https://www.rakuten.co.jp/} for constructing KGs, and obtain the general user embeddings based on the KGs.
The procedure of constructing KGs is as follows: 
1) collect \textit{seed users}, who are picked up based on purchase history from Rakuten Ichiba users. 
2) collect all the items purchased by them. 
3) extract demography features of the seed users and attributes of the items.
4) construct a KG: connect users and items with purchase history; attributes are extracted as entities and they are connected to users and items.
After creating the KG, we apply the framework of KQGC to it, then we finally obtain the general user embedding for the seed users.
\subsubsection{Training}
As our goal here is to obtain user embeddings from the KQGC, there is no decoder in the downstream task. Therefore, we employ the following unsupervised loss function for training KQGC.
\begin{align}
    \mathcal{L}_{\textrm{CF}} &= \sum_{(\mathcal{U}, \mathcal{I})\in\mathcal{P}}\sum_{(\mathcal{U}, \mathcal{I}')\in\mathcal{P}^{-1}}\left[\gamma + f(\bm{h}_{\mathcal{U}}, \bm{h}_{\mathcal{I}}) - f(\bm{h}_{\mathcal{U}}, \bm{h}_{\mathcal{I}'})\right]_+,\\
    \label{eq:transE_CF}
    f(\bm{h}_{\mathcal{U}}, \bm{h}_{\mathcal{I}}) &= \|\bm{h}_{\mathcal{U}} + \bm{h}_{\textrm{purchase}} - \bm{h}_{\mathcal{I}} \|_{1,2},
\end{align}
where $(\mathcal{U}, \mathcal{I})\in\mathcal{P}$ is the positive pairs between users and items with purchase relation, $(\mathcal{U}, \mathcal{I}')\in\mathcal{P}^{-1}$ is the negative one, $f(\bm{h}_{\mathcal{U}}, \bm{h}_{\mathcal{I}})$ is the score function especially for user-item interaction, $\bm{h}_{\mathcal{U}}$ and $\bm{h}_{\mathcal{I}}$ are user and item embeddings, $\bm{h}_{\textrm{purchase}}$ is purchase-relation embeddings. $\mathcal{I}'$ denotes negative items, which are generated by replacing the positive item with an item randomly chosen in the positive pairs. Here we employ L1 norm for Eq. (\ref{eq:transE_CF}).

\section{Experiments}
In this section, we conduct experiments to evaluate the performance of our model using the same metric (PR-AUC) which is being used in AIris TP for evaluation. The KQGC model has been implemented to the Rakuten AIris TP to provide effective target prospecting. We first explain the AIris TP system overview and experimental settings. Then, we evaluate the effectiveness of our model with five brand datasets (the name of the brands have been kept anonymous ) by comparing it with the baseline model. Finally, we briefly discuss the effectiveness of the model in the real target prospecting scenario.

\subsection{Overview of the Rakuten AIris TP}

\begin{figure}[htbp]
    \begin{center}
    \includegraphics[width=\linewidth]{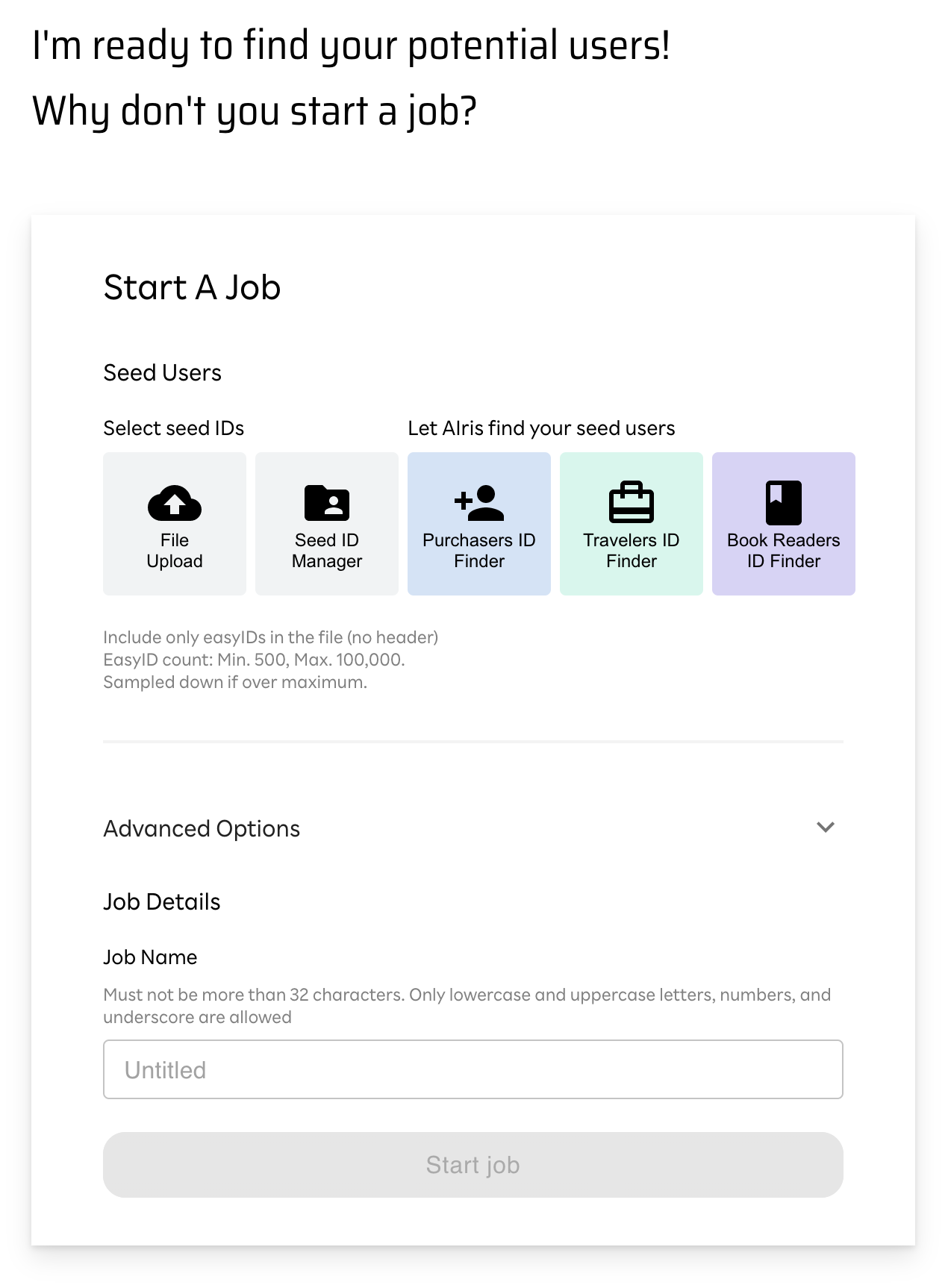}
    \caption{Rakuten AIris TP Framework for marketers and clients.}
    \label{fig:airis_tp}
    \end{center}
\end{figure}

In this section, we briefly describe the Rakuten Airis TP framework. Figure \ref{fig:airis_tp} shows the dashboard of AIris TP for clients. A client/marketer can upload a file of users (to find out similar users) or she can use “Seed ID Manager ” to select appropriate seed users for the task or retrieve the previous purchasers of a product/brand using “Purchaser ID Finder”. There are two more options for travel (Travelers ID Finder) and books (Book Readers ID Finder) related targeting. After uploading the seed users using one of the options mentioned above, clients can start a job for finding lookalike customers for their business purpose.
\subsection{Experimental set-up}
\label{subsec:experimental_setup}
\subsubsection*{\textbf{Datasets}} 
In AIris TP, for a specific product/brand we pick up the users who already bought that product/brand in previous months and then AIris TP extracts potential customers for a product/brand who have high probability to buy the product/brand in next month. We adopt five cosmetic brands as the AIris clients in this experiment and evaluate models based on whether AIris can extract accurate potential customers for each of the client from given users, with the features obtained from the models. \par
To treat this problem as binary classification, we arrange both positive users (label 1) and negative users (label 0) for each client, in the datasets of TP. Positive users are those who purchased at least one item of the client in a certain time-window, i.e., positive users correspond to the accurate potential customers in this evaluation. On the other hand, negative users are those who purchased no item in the same time-window as positive users. Ratio between positive and negative users is set to around 1 : 3 for each client. In terms of data split, we collect train/validation and test users in different time-windows to remove possible data leakage. 20K users are collected in two months from Jul. 1st, 2021 to Oct. 31st, 2021 as training and validation data sets of each client, and they are randomly divided into 6 (training) : 4 (validation). 5K test users for each client are collected in Sep. 1st, 2021 - Sep. 31st, 2021 (one month). If the number of positive users for a client in the time-window is less than 20K or 5K, the number is used instead of 20K or 5K. Purchase behavior for constructing KG is collected in Feb. 1st, 2021 - Jul. 31st, 2021, following the way described in \ref{subsubsec:how_to_construct_graph}.
 Statics of the datasets of TP and the KG used for training KQGC are shown in table \ref{table:airis_datasets} and table \ref{table:kg_datasets}, respectively. 

\begin{table*}[hbtp]
    \caption{Statics of the AIrist datasets}
    \label{table:airis_datasets}
    \centering
    \begin{tabular}{ll ccccc}
      \hline
      &&Brand A &Brand B &Brand C &Brand D &Brand E \\
      \hline\hline
      \multirow{3}{*}{Training} &\# positive users  &3784  &3768  &3719  &1668 &3753 \\
                             &\# negative users  &11216 &11232 &11281 &5504 &11247 \\
                             &\# total users     &15000 &15000 &15000 &6672 &15000 \\
      \hline
      \multirow{3}{*}{Validation} &\# positive users  &1255 &1241 &1285 &213 &1206 \\
                                  &\# negative users  &3745 &3759 &3715 &639 &3794 \\
                                  &\# total users     &5000 &5000 &5000 &852 &5000 \\
      \hline
      \multirow{3}{*}{Test} &\# positive users  &1285 &1238 &1264 &1100 &1233 \\
                            &\# negative users  &3715 &3715 &3736 &3300 &3767 \\
                            &\# total users     &5000 &5000 &5000 &4400 &5000 \\ 
      \hline
    \end{tabular}
\end{table*}

\begin{table}[hbtp]
    \caption{Statics of the KG for training KQGC}
    \label{table:kg_datasets}
    \centering
    \begin{tabular}{cccc}
      \hline
      \# users &\# items &\# edges &\# relations \\
      \hline\hline
      85,402  &1,302,877  &2,360,012  &1 (purchase) \\
      \hline
    \end{tabular}
\end{table}

\subsubsection*{\textbf{Baselines}} To demonstrate the effectiveness, we compare
our proposed model with the current model and on production and other KGE models.

\begin{itemize}
\item \textbf{Baseline (AIris TP)}: This is the current model on production (see Fig. \ref{fig:airis_tp}). It is based on XGBoost model. This model uses five different features (note that customer true identity is totally hidden to the dataset and model):
 \\
 \\
 \textit{Demographic Features}: Demographic features such as age, gender, region.\\
 \textit{Point Summary}: Rakuten users can gain points while buying different products/services. This feature exhibits point status such as current available points. \\
 \textit{Point Feature}: Transaction of points such as acquired/used points from online/offline shops/ merchants. \\
 \textit{Genre level Purchase History}: Like other E-commerce companies Rakuten group maintains "genre" hierarchy. In this feature we capture the shopping trends in popular genres. \\
 
 \item\textbf{TransE}: Only TransE generated vectors has been employed here for AIris TP user prediction as baseline.\\
\end{itemize}

\subsubsection*{\textbf{Evaluation}}
We evaluate PR-AUC of AIris with user embeddings (input features) obtained from each model. Three types of input features are compared in this experiment: baseline features, KGE of TransE and KQGC.
The baseline features are user attributes including demography, genre preference, whose dimension is 157.
In terms of KGE of TransE and KQGC, we concatenate each KGE with the baseline features, and then input it to AIris, i.e., the dimension of the input features is 257 when that of KGE is 100.
\subsubsection*{\textbf{Model settings}}
The models are implemented in PyTorch and Deep Graph Library \cite{DGL}. The training and inference run on a single GPU 'NVIDIA Tesla V100'. Hyperparamters of TransE are as follows. The number of epochs is 10,000, batch size is 10,000, learning rate is 0.001, dimension of embeddings is 100, ratio between positive and negative triples is 1 : 1, and margin $\gamma$ is 1.0. Hyperparameters of KQGC are as follows. The number of GNN layers is 1, hidden and output layers is 100, KGE of TransE in 5000 epochs is used as input of KQGC, ratio between positive and negative pairs is 1 : 3, graph sampling where 10 neighbor nodes are randomly sampled is used only in training phase, and others are the same as TransE. Trainable parameters are initialized with Xavier normalization.
\subsection{Results}
\begin{figure}[htbp]
    \begin{center}
    \includegraphics[]{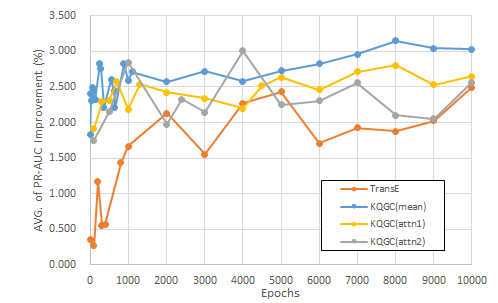}
    \caption{Evolution of the average of PR-AUC improvement rate with epochs}
    \label{fig:evo_prauc}
    \end{center}
\end{figure}

Our goal is to improve PR-AUC for various clients.
Hence we hereafter emphasize the average of the PR-AUCs for all the clients. Note that user embeddings and baseline features are shared among all the clients, on the other hand, AIris is optimized for each client independently. The best PR-AUCs over 10000 epochs are shown in table \ref{table:results}. In table \ref{table:results}, the values in parentheses are improvement rate (\%) compared to the baseline and the other values are PR-AUCs. The highest PR-AUC among different input features is marked in bold for each column. Table \ref{table:results} shows that KQGC outperforms both the baseline and the original TransE except for Brand D, and KQGC(mean) reaches the best PR-AUC for the average of all the clients.
Fig. \ref{fig:evo_prauc} shows evolution of PR-AUC of the averaged improvement rate with epochs. From Fig. \ref{fig:evo_prauc}, we can observe that KQGC(mean and attn1) achieves more stable performance than others. That mean aggregator's performance is the best and attn2 aggregator's one is unstable possibly because the embedding module and the prediction one are separated. The attention mechanism of KQGC are automatically designed by the loss function of pre-training (Eq.\ref{eq:transE_CF}), therefore it is difficult to construct an attention mechanism that is suitable for TP. 

\begin{table*}[hbtp]
    \caption{Improvement rate compared to the baseline model in PR-AUC }
    \label{table:results}
    \centering
    \begin{tabular}{lcccccc}
      \hline
      Input features &Brand A &Brand B &Brand C &Brand D &Brand E &AVG. \\
      \hline\hline
      Baseline              &0.693 &0.678 &0.593 &0.749 &0.732 &0.689 \\
      TransE       &0.706 (+1.83) &0.699 (+3.13) &0.604 (+1.90) &\textbf{0.786} (+4.95) &0.737 (+0.97) &0.706 (+2.50)\\
      KQGC(mean)  &\textbf{0.712} (+2.70) &0.703 (+3.62) &0.607 (+2.46) &0.784 (+4.65) &\textbf{0.749} (+2.32) &\textbf{0.711} (+3.15)\\
      KQGC(attn1) &0.708 (+2.11) &\textbf{0.705} (+3.88) &0.607 (+2.52) &0.781 (+4.26) &0.741 (+1.23) &0.708 (+2.80)\\
      KQGC(attn2) &0.710 (+2.46) &0.702 (+3.55) &\textbf{0.612} (+3.27) &0.784 (+4.67) &0.740 (+1.09) &0.710 (+3.00)\\
      \hline
    \end{tabular}
\end{table*}

\subsection{Model Deployment}
The AIris TP is a well-known Rakuten target prospecting framework.  Rakuten group has started this service from the year of 2018. Periodically we evaluate the deployed model’s performance and improve the model. The proposed KQGC is in the testing phase for deployment on our AIris TP framework. The major advantage of AIris TP is: it allows the marketers/clients to perform the target prospecting without training the entire model from scratch when we have a new task (job), e.g., find the potential users with similar behavior patterns to new brand’s existing users. The tentative customer conversion rate (CVR) is expected to go much higher in comparison with the previous standard solution. 

\section{Conclusion}
In this paper, we have introduced a novel model for recommender systems that leverages both KGE and graph convolution. KQGC smooths knowledge graph embedding (KGE) obtained from translation-based models with the aggregation of neighbor knowledge queries, and helps adjacent entities have similar embeddings. Empirical result on Rakuten group's real data shows that KQGC outperforms current AIris TP baseline models including TransE, which is the original translation-based model, and we also observed smoothing KGE is effective for improving performance on recommendation tasks.\par
In future work, we will explore the following three things. 
(1) Deeper analysis of our model; In this work, our model employs one convolutional layer, and it is evaluated on only our private dataset. To confirm the effectiveness of our model more strongly, we will evaluate our model with more than two layers, and compare it with several SoTA models on various public datasets.
(2) Incorporation of side information; continuous attributes such as image- or text-embeddings are difficult to be extracted as entities on a KG. Hence we need to investigate how we can effectively combine KGE with those continuous attributes as inputs of KQGC.
(3) Generality of our model; Our ultimate goal is to apply user embeddings obtained from KQGC to various services. As this paper focuses on prospect recommendation task, we will evaluate the user embeddings on different services and investigate the generality as our next research direction.

\bibliographystyle{ACM-Reference-Format}
\bibliography{main}


\end{document}